\documentclass[aps,prc,twocolumn,superscriptaddress]{revtex4-1}

\usepackage{amsmath,amssymb,graphicx}
\usepackage{url}
\usepackage[colorlinks,urlcolor=blue]{hyperref}

\begin{document}

%\preprint{}

\title{Search for the Elusive Jet-Induced Diffusion Wake in $Z/\gamma$-Jets with 2D Jet Tomography  in High-Energy Heavy-Ion Collisions}



\author{Wei Chen}
%\email[]{cw1987@mails.ccnu.edu.cn}
%\thanks{}
\affiliation{School of Nuclear Science and Technology, University of Chinese
Academy of Sciences, Beijing 100049, China}
%\affiliation{Key Laboratory of Quark and Lepton Physics (MOE) and Institute of Particle Physics,Central China Normal University, Wuhan 430079, China}

\author{Zhong Yang}
\affiliation{Key Laboratory of Quark and Lepton Physics (MOE) \& Institute of Particle Physics,Central China Normal University, Wuhan 430079, China}

\author{Yayun He}
\affiliation{Key Laboratory of Quark and Lepton Physics (MOE) \& Institute of Particle Physics,Central China Normal University, Wuhan 430079, China}

\author{Weiyao Ke}
\affiliation{Physics Department, University of California, Berkeley, California 94720, USA}
\affiliation{Nuclear Science Division MS 70R0319, Lawrence Berkeley National Laboratory, Berkeley, California 94720, USA}

\author{Long-Gang Pang}
\affiliation{Key Laboratory of Quark and Lepton Physics (MOE) \& Institute of Particle Physics,Central China Normal University, Wuhan 430079, China}

\author{Xin-Nian Wang}
%\email[]{xnwang@lbl.gov}
\affiliation{Key Laboratory of Quark and Lepton Physics (MOE) \& Institute of Particle Physics,Central China Normal University, Wuhan 430079, China}
\affiliation{Physics Department, University of California, Berkeley, California 94720, USA}
\affiliation{Nuclear Science Division MS 70R0319, Lawrence Berkeley National Laboratory, Berkeley, California 94720, USA}
%\thanks{Current address}
%\affiliation{Physics Department, University of California, Berkeley, California 94720}


\date{January 13, 2021}

\begin{abstract}
Diffusion wake is an unambiguous part of the jet-induced medium response in high-energy heavy-ion collisions that leads to a depletion of soft hadrons in the opposite direction of the jet propagation. New experimental data on $Z$-hadron correlation in Pb+Pb collisions at the Large Hadron Collider show, however, an enhancement of soft hadrons in the direction of both the $Z$ and the jet. Using a coupled linear Boltzmann transport and hydro model, we demonstrate that medium modification of partons from the initial multiple parton interaction (MPI) gives rise to a soft hadron enhancement that is uniform in azimuthal angle while jet-induced medium response and soft gluon radiation dominate the enhancement in the jet direction. After subtraction of the contributions from MPI with a mixed-event procedure, the diffusion wake becomes visible in the near-side $Z$-hadron correlation. We further employ the longitudinal and transverse gradient jet tomography for the first time to localize the initial jet production positions in $Z/\gamma$-jet events in which the effect of the diffusion wake is apparent in $Z/\gamma$-hadron correlation even without the subtraction of MPI. 

% And it also shows the medium modification of particle ratio (p/$\pi$, $\Lambda$/$k_s^0$) in the reconstructed jet in the small $p_T$ range.
\end{abstract}

%\keywords{}

\maketitle


\noindent{\it 1. Introduction}: Since the discovery at the Relativistic Heavy-ion Collider (RHIC) \cite{Adcox:2001jp,Adler:2002xw,Gyulassy:2004zy,Wang:2004dn} and the confirmation at the Large Hadron Collider (LHC) \cite{Aad:2010bu,Aamodt:2010jd,Chatrchyan:2011sx,Majumder:2010qh,Muller:2012zq,Qin:2015srf}, jet quenching or the suppression of large transverse momentum jets and  hadrons due to jet-medium interaction has enabled one to glean properties of the strongly interacting quark-gluon plasma (QGP) formed in high-energy heavy-ion collisions. For example, the extracted jet transport coefficients in the QGP in central heavy-ion collisions at RHIC and LHC
%by the JETSCAPE Collaboration 
\cite{Burke:2013yra} are two orders of magnitude higher than that in a cold nucleus \cite{Wang:2009qb}. Such strong jet-medium interaction should also lead to jet-induced medium response in the form of Mach-cone-like excitations \cite{CasalderreySolana:2004qm,Stoecker:2004qu,Ruppert:2005uz,Gubser:2007ga,Gubser:2007ga,Qin:2009uh,Li:2010ts}.
%due to the propagation of the energy-momentum lost by jets to the medium.
Study of such jet-induced medium excitations can provide further information about the QGP such as the bulk transport properties and the equation of state \cite{Bouras:2012mh,Ayala:2016pvm,Yan:2017rku,Cao:2020wlm}.

There are many consequences of the jet-induced medium excitation \cite{Cao:2020wlm,Tachibana:2020mtb}, from the broadening of the jet profile at large radius \cite{Tachibana:2014lja,Tachibana:2017syd,Luo:2018pto} to cone-size dependence of the jet suppression \cite{He:2018xjv,Pablos:2019ngg}, modification of the jet fragmentation function \cite{Chen:2020tbl,Casalderrey-Solana:2016jvj}, jet-hadron and $Z/\gamma$-hadron correlations \cite{Ma:2010dv,Betz:2010qh,Chen:2017zte,Yan:2017rku}. Isolation and detailed study of the medium excitation are therefore fundamental to the understanding of these phenomena and crucial for using them to extract bulk properties of the QGP. 

The jet-induced medium response consists of the wake front and the diffusion wake. Microscopically, each jet-medium interaction kicks the medium parton into a recoil particle and leaves behind a “particle-hole”. Further transport of the recoil particles forms the wake front, while the diffusion of the “particle-holes” leads to the diffusion wake \cite{He:2015pra}. The wake front is known to contribute to an enhancement of soft hadrons with an energy scale of the medium temperature $\omega \sim T$ within the jet cone \cite{Casalderrey-Solana:2015vaa,Chen:2017zte,Chen:2020tbl,Casalderrey-Solana:2020rsj}. However, it is difficult to separate this contribution from medium-induced soft gluon radiations whose energy scale is also around $\omega_g\sim \hat q \lambda^2 \sim T$ \cite{Blaizot:2013hx,Schlichting:2020lef,Ke:2020clc} when their typical formation time $\tau_f\sim \sqrt{2\omega_g/\hat q}$ is limited by the mean-free-path $\lambda\sim 1/T$. Here $\hat q\sim T^3$ is the jet transport coefficient in the QGP. The diffusion wake, on the other hand, leads to a depletion of soft hadrons in the back direction of the jet and, therefore, is an unambiguous signal of the jet-induced medium excitation without any similar competing effect.
Such a depletion is indeed observed in the calculated $\gamma$-hadron correlation at RHIC within the coupled linear Boltzmann transport (CoLBT) hydro model \cite{Chen:2017zte} and di-jet correlation with a large rapidity gap at LHC within the hybrid strong/weak coupling model \cite{Pablos:2019ngg}. However, new experimental data \cite{Sirunyan:2021jkr} show an enhanced soft $Z$-hadron yield in the $Z$ direction in Pb+Pb collisions at $\sqrt{s}_{\rm NN}=5.02$ TeV instead of an expected depletion.

In this Letter we will present our study of $Z/\gamma$-hadron correlation in Pb+Pb collisions at LHC within the CoLBT-hydro model. We will examine the effect of the initial multiple parton interaction (MPI) which is negligible at RHIC but becomes sizable at LHC. We will show that medium modification of MPI partons leads to an enhanced soft $Z/\gamma$-hadron correlation which is uniform in the azimuthal angle and offsets the effect of the diffusion wake. We will devise a mixed-event procedure to subtract the contribution from MPI so that the diffusion wake becomes visible in the $Z/\gamma$-hadron correlation. We will further apply the longitudinal and transverse gradient jet tomography for the first time to select events in which the initial positions of $Z/\gamma$-jet production are localized and the effect of diffusion wake is clear even without the subtraction of MPI.

\noindent{\it 2. $Z/\gamma$-jet modification in heavy-ion collisions}:
We will use the CoLBT-hydro model \cite{Chen:2017zte} to study $Z/\gamma$-jet production and jet-induced medium response in high-energy heavy-ion collisions. CoLBT-hydro couples jet propagation within the linear Boltzmann transport (LBT) model \cite{He:2015pra} to the event-by-event (3+1)D CCNU-LBNL viscous (CLVisc) hydrodynamic model \cite{Pang:2018zzo} in real time through a source term from the energy-momentum lost to the medium by jet shower and recoil partons. The LBT model \cite{He:2015pra} is based on the Boltzmann equation for both jet shower and recoil partons with pQCD leading-order elastic scattering and induced gluon radiation according to the high-twist approach \cite{Guo:2000nz,Wang:2001ifa,Zhang:2003yn,Zhang:2003wk}. The final hadron spectra from CoLBT-hydro include contributions from the hadronization of hard partons within a parton recombination model \cite{Han:2016uhh} and jet-induced hydro response via Cooper-Frye freeze-out.  A freeze-out temperature $T_f=137$ MeV and specific shear viscosity $\eta/s=0.08$ together with the s95p parameterization of the equation of state \cite{Huovinen:2009yb} and AMPT \cite{Lin:2004en} or Trento \cite{Bernhard:2016tnd} initial conditions are used in CLVisc which can reproduce experimental data on bulk hadron spectra and anisotropic flows at both RHIC and LHC energies \cite{Pang:2018zzo}. For more detailed descriptions of LBT and CoLBT-hydro model we refer readers to Refs.~\cite{He:2015pra,Cao:2016gvr,Cao:2017hhk,He:2018xjv,Luo:2018pto,Zhang:2018urd} and \cite{Chen:2017zte,Chen:2020tbl}.

%For simulations of $Z/\gamma$-jet production and propagation in heavy-ion collisions at the LHC within the CoLBT-hydro model, 
\begin{figure}
\centerline{\includegraphics[scale=0.34]{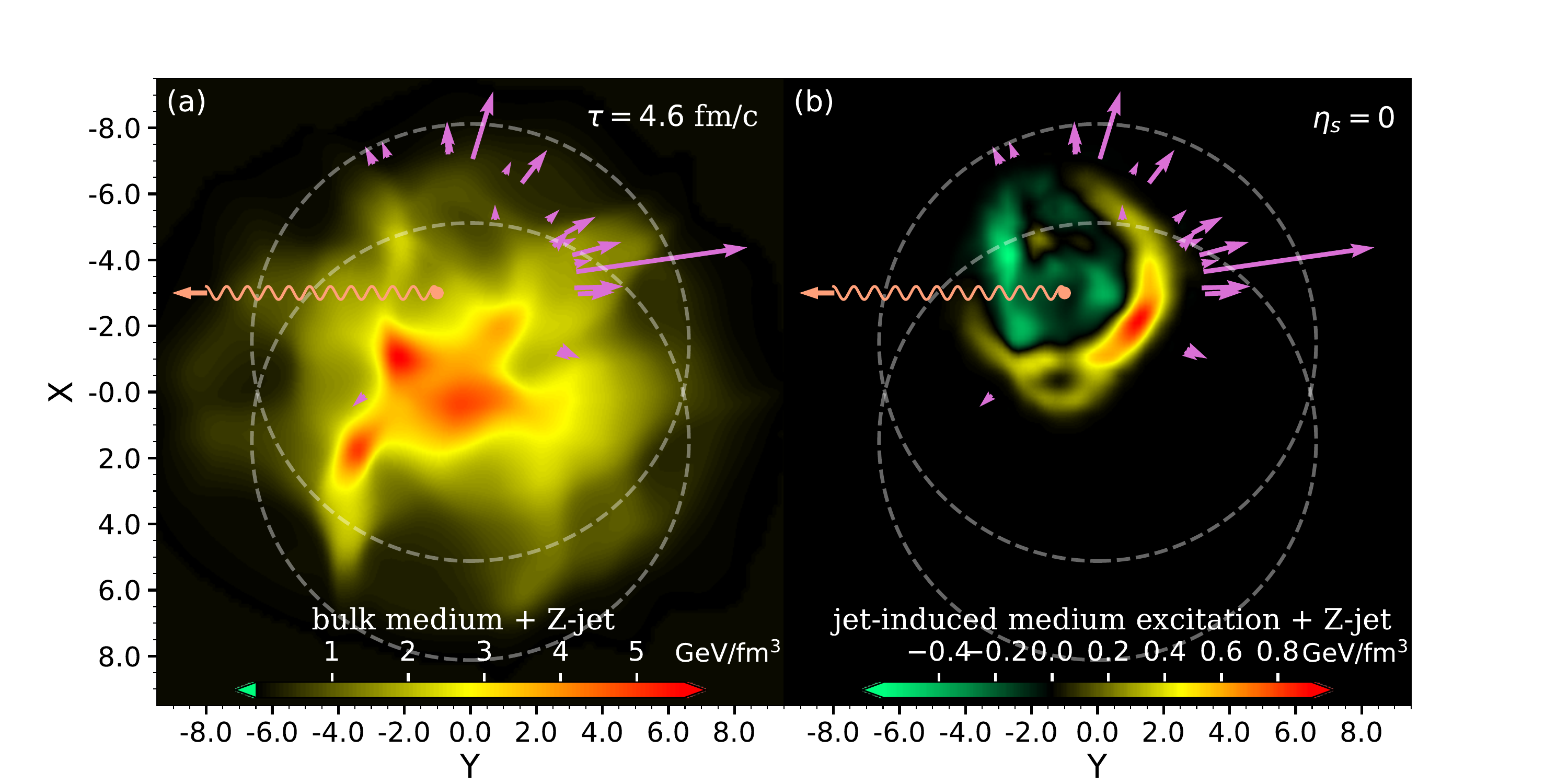}}
\caption{ A snapshot of (a) the total and (b) jet-induced energy density distribution in the transverse plane of a semi-central Pb+Pb collision at $\sqrt{s}_{\rm NN}=5.02$ TeV from CoLBT-hydro simulations with a $Z$-jet at the spatial rapidity $\eta_s=0$ and proper time $\tau=4.6$ fm/$c$. Straight (wavy) lines represent the transverse momenta of partons ($Z$ boson) and dashed circles represent the two colliding nuclei.}
\label{fig-machcone}
\end{figure}

PYTHIA8 \cite{Sjostrand:2007gs} is used to generate initial $Z/\gamma$-jet configurations. The isospin dependence of the parton distribution functions (PDF) in a nucleus is considered but other cold nuclear modification of the PDF is neglected since it has negligible effects on jet and hadron spectra per $Z/\gamma$ trigger. The initial transverse positions of  $Z/\gamma$-jets are sampled according to the binary nucleon-nucleon collisions within the HIJING \cite{Wang:1991hta,Gyulassy:1994ew} and the AMPT model \cite{Lin:2004en} which also provides initial conditions for CLVisc hydro simulations.  Partons from jet showers as well as from MPI's associated with the $Z/\gamma$ trigger are allowed to propagate through the QGP and generate medium response according to the CoLBT-hydro model.
%The interaction between these initial partons and the QGP medium starts after the initial time $\tau_0$ for the hydrodynamics or their formation time whichever is later. 

To illustrate the jet-induced medium response in A+A collisions, we show in Fig.~\ref{fig-machcone} (a) the transverse distribution of the energy density at $\tau=4.6$ fm/$c$ in a semi-central Pb+Pb collision with a $Z$-jet at $\sqrt{s}_{\rm NN}=5.02$ TeV. The transverse momenta of hard partons are indicated by arrowed lines and the direction of the $Z$ trigger by the wavy line.  After subtracting the energy density from the same hydro event without the $Z$-jet, we obtain the energy density distribution of the jet-induced medium excitation, which has a wake front (positive) and the diffusion wake (negative energy density) as shown in  Fig. 1 (b).
The $Z$-jet shown in Fig.~\ref{fig-machcone} is produced off-center and propagates tangentially through the bulk medium. The jet-induced medium response and hard partons are therefore both distorted by the density gradient and radial flow. Such asymmetrical distortion will show up in the $Z$-hadron azimuthal correlation and will provide an unambiguous signal of the diffusion wake.

\begin{figure}
\includegraphics[scale=0.55]{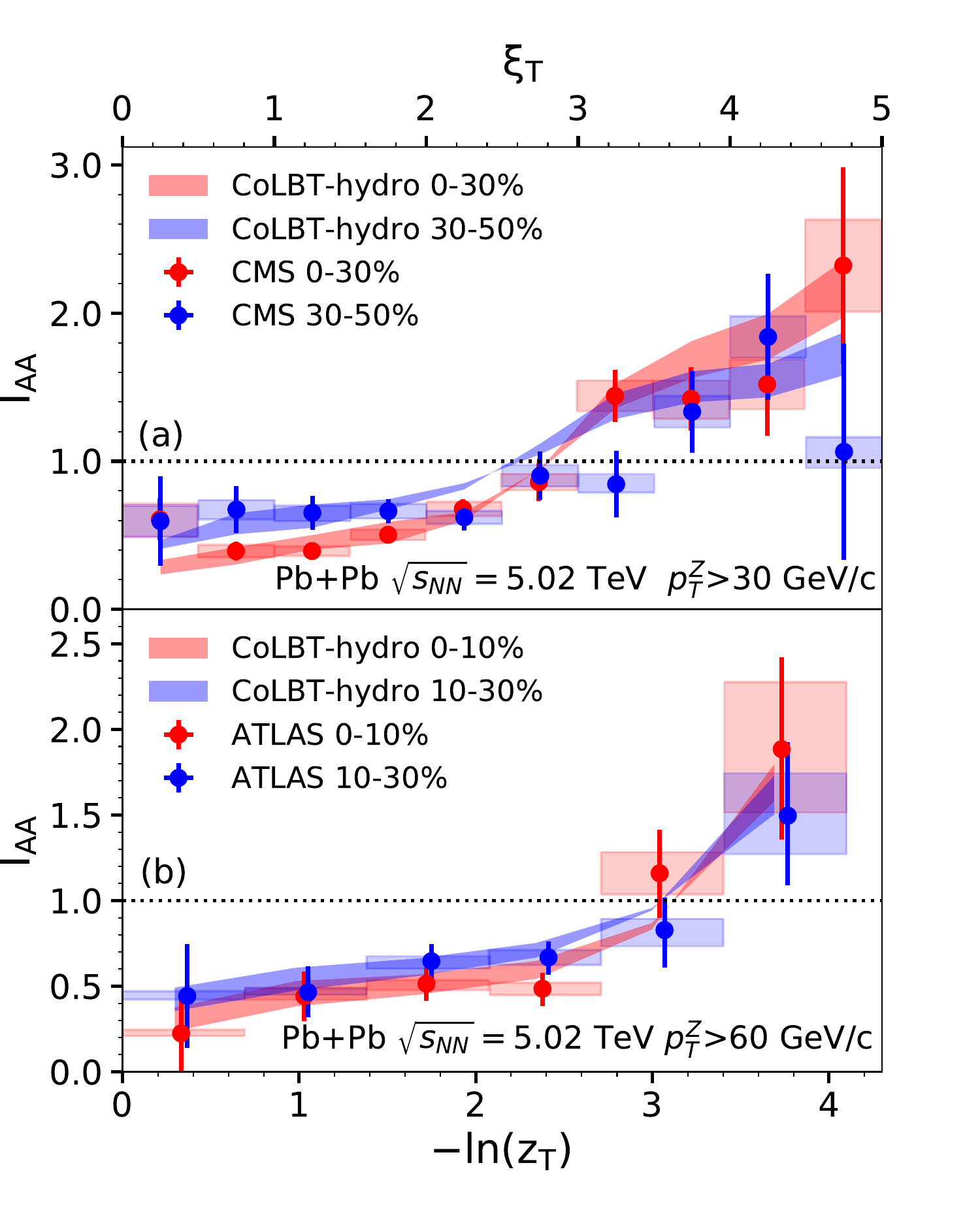}
\caption{Modification factors $I_{AA}$ for $Z$-triggered yield of charged hadrons as a function of $\xi_T$ or $\ln(1/z_T)$ in semi-central Pb+Pb collisions at $\sqrt{s}_{\rm NN}=5.02$ TeV compared to (a) CMS data \cite{Sirunyan:2021jkr} for $p_T^Z>30$ GeV/$c$, $|\Delta\phi^{hZ}|>7\pi/8$ and (b) ATLAS data \cite{ATLAS:2020wmg} for $p_T^Z>60$ GeV/c, $|\Delta\phi^{hZ}|>3\pi/4$.}
\label{fig-Iaa}
\end{figure}

In CoLBT-hydro,  hadron spectra associated with $Z/\gamma$ production have contributions from both the hadronization of hard partons and the jet-induced hydro response which is calculated from the bulk hadron spectra with $Z/\gamma$ trigger  minus that from the same hydro events but without $Z/\gamma$.
Shown in Fig.~\ref{fig-Iaa}(a) (Fig.~\ref{fig-Iaa}(b)) are the modification factors,
    $I_{\rm AA}=dN_{\rm AA}^{hZ}/dN_{\rm pp}^{hZ}$,
for  charged hadron spectra associated with a $Z$ trigger in 0-30\% (red) and 30-50\% (blue) (0-10\% and 10-30\%) Pb+Pb collisions at $\sqrt{s}_{\rm NN}=5.02$ TeV as a function of $\xi_T=\ln(-{|\vec p_T^Z|}^2/\vec p_T^h\cdot \vec p_T^Z)$ ($-\ln z_T=\ln[p_T^Z/p_T^h]$) as compared to the CMS \cite{Sirunyan:2021jkr} (ATLAS \cite{ATLAS:2020wmg}) data. The transverse momentum of the $Z$ is $p_T^Z> 30$ (60) GeV/$c$, hadron $p_T^h>1$ GeV/$c$ and their azimuthal angles are restricted to $|\Delta\phi^{hZ}|=|\phi^h-\phi^Z|>7\pi/8$ $(3\pi/4)$. The effective strong coupling constant $\alpha_s=0.28\pm 0.02$ in CoLBT-hydro is fitted to the CMS data on $I_{\rm AA}$ in 0-30\%  Pb+Pb collisions. The bands in the CoLBT-hydro results correspond to the variation of the effective strong coupling constant $\alpha_s$ in the model within 95\% credible region of the Bayesian fitting probability.

We can see that CoLBT-hydro describes well within the experimental errors both the suppression of leading hadrons at small $\xi_T$ (large $z_T$) due to parton energy loss and soft hadron enhancement at large $\xi_T$ (small $z_T$) due to jet-induced medium excitation as well as medium-induced gluon radiation. Since soft hadrons from medium induced gluons have the same energy scale of the local temperature, due to the constraints imposed by the formation time \cite{Blaizot:2013hx,Schlichting:2020lef,Ke:2020clc}, as hadrons from jet-induced medium response, it is difficult to separate these two contributions experimentally.  Note that $Z$-triggered hadron spectra in CMS data \cite{Sirunyan:2021jkr} and CoLBT-hydro results contain contributions from MPI which are subtracted in the ATLAS data \cite{ATLAS:2020wmg}. 
%We discuss this contribution in detail next.

\begin{figure}
\includegraphics[scale=0.54]{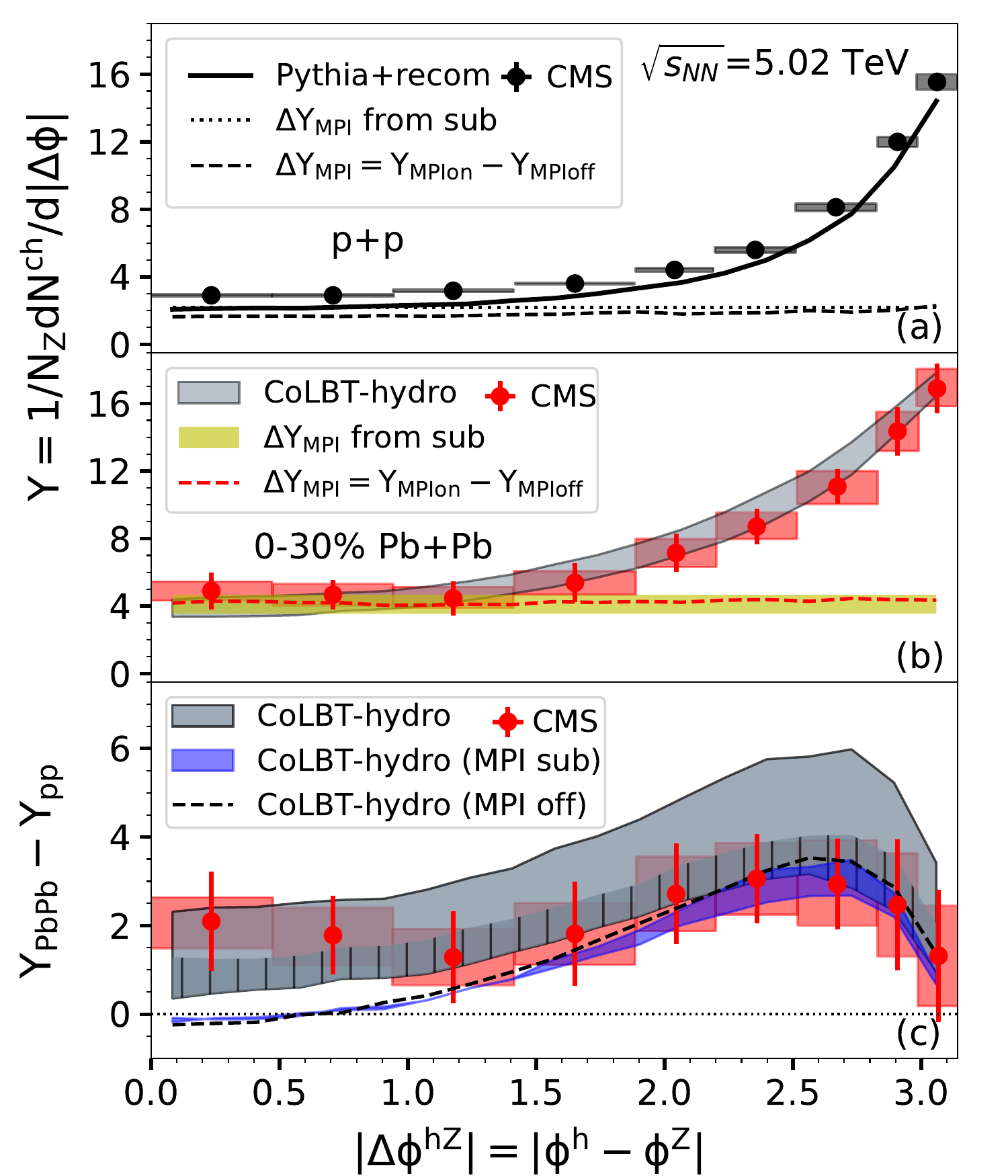}
\caption{ Charged hadron yields per $Z$-trigger as a function of $|\Delta \phi^{hZ}|$ in (a) p+p (solid line) and (b) 0-30\% Pb+Pb collisions (grey band) and (c) their difference (grey band) at $\sqrt{s}_{\rm NN}=5.02$ TeV as compared to CMS data \cite{Sirunyan:2021jkr} for $p_T^Z>30$ GeV/$c$ and $p_T^h>1$ GeV/$c$. Dotted (dashed) line in (a) and yellow band (dashed line) in (b) are the MPI contributions from the subtraction procedure (by subtracting the yield in events with MPI off) in CoLBT-hydro simulations. The blue band (dashed line)  in (c) is the yield difference after subtraction of MPI (with MPI off). See text for detailed explanations.}
\label{fig-dphi}
\end{figure}

\noindent{\it 3. $Z/\gamma$-hadron correlation and MPI}:
To examine the medium modification of the $Z/\gamma$-triggered hadron spectra and isolate soft hadrons from the jet-induced medium response and the diffusion wake in particular, we compute the $Z/\gamma$-hadron correlation in azimuthal angle.  Shown in Fig.~\ref{fig-dphi} are (a) charged hadron yields per $Z$-trigger in p+p (solid line) and (b) 0-30\% Pb+Pb (grey band) collisions and (c) their difference (grey band) as a function of the azimuthal angle $|\Delta\phi^{hZ}|$ as compared to the CMS data \cite{Sirunyan:2021jkr} for $p_T^Z>30$ GeV/$c$ and $p_T^h>1$ GeV/$c$ at $\sqrt{s}_{\rm NN}=5.02$ TeV. One can see there are both an enhancement and a broadening of the peak in the jet direction in Pb+Pb as compared to p+p collisions. The hadron yield in the $Z$ direction is also sizable which is further enhanced in Pb+Pb collisions. This is in contrary to the expectation of a depletion due to jet-induced diffusion wake. We find that hadrons in the $Z$ direction comes mainly from MPI associated with a triggered hard process as have been investigated in p+p collisions \cite{Acosta:2004wqa,Aad:2010fh,Khachatryan:2010pv,ALICE:2011ac,Adam:2019xpp}. MPI's are independent minijet production in PYTHIA8 \cite{Sjostrand:1987su,Wang:1991hta}. The corresponding $Z$-hadron correlation is therefore uniform in the azimuthal angle. Interaction of these minijets with the medium will lead to an enhancement of soft hadrons and suppression of large $p_T$ hadrons from MPI's. This explains the enhanced soft hadron yield ($p_T^h\sim 1-2$ GeV/$c$) in the $Z$ direction in Pb+Pb collisions as we see in both the CMS data and CoLBT-hydro results. Shown in Figs.~\ref{fig-dphi}(a) and (b) as dashed lines are hadron yields from MPI in p+p and Pb+Pb  collisions by subtracting $Z$-hadron yields with and without MPI in CoLBT-hydro simulations. 
%Note that contributions from $Z$-($n>1$)-jets in the $Z$ direction are negligible as compared to MPI \cite{Luo:2018pto,Zhang:2018urd}. 

The p+p results in Fig.~\ref{fig-dphi}(a) are from PYTHIA8 with the same recombination model~\cite{Han:2016uhh} for hadronization as in CoLBT-hydro. They are systematically smaller than the CMS data and the default PYTHIA8 result. We consider this discrepancy, possibly due to the neglect of color correlation among MPI minijets in the recombination model, as a systematic error in our model calculations and is added to the error band (hatched area) in the medium modification of the $Z$-hadron correlation in Fig.~\ref{fig-dphi}(c). Since the minijet cross section at the RHIC energy is about a factor of 10 smaller~\cite{Wang:1991hta} than at LHC, the effect of MPI at the RHIC energy is much smaller \cite{Adam:2019xpp}, making it easier to isolate the effect of diffusion wake in $\gamma$-hadron correlations

To search for signals of the jet-induced diffusion wake, we devise a procedure to subtract the MPI's contribution to the $Z/\gamma$-hadron correlation. We first calculate the uniform correlation between $Z/\gamma$ in one event and hadrons from another similar $Z/\gamma$-jet event. We assume the effect of the diffusion wake on the total $Z/\gamma$-hadron yield in the mixed events is negligible. Contributions from jets to the $Z/\gamma$-hadron correlation in these mixed events, which are assumed to be the same as the integrated $Z/\gamma$-hadron yield within an angle $|\Delta\phi^{hZ}|>1$ in $Z/\gamma$-jet events in addition to the MPI background, can be subtracted approximately to obtain the MPI's contribution,
\begin{equation}
\frac{dN^{hZ}_{\rm MPI}}{d\phi}\approx \frac{dN^{hZ}_{\rm mix}}{d\phi}-\int_1^\pi \frac{d\phi}{\pi} \left(\frac{dN^{hZ}}{d\phi}-\frac{dN^{hZ}}{d\phi}|_{\phi=1}\right).
\end{equation}
These MPI contributions to the $Z/\gamma$-hadron correlation are shown as the dotted line for p+p in Fig.~\ref{fig-dphi}(a) and the yellow band for Pb+Pb collisions in Fig.~\ref{fig-dphi}(b). After subtracting the above MPI contributions, one can obtain the true medium modification of the $Z$-hadron correlation due to jet quenching and the jet-induced medium response, shown as the blue band in Fig.~\ref{fig-dphi} (c). Comparison to the medium modification without MPI (dashed) shows that the subtraction procedure works well. The $Z$-hadron correlation in the $Z$ direction indeed becomes slightly negative after the subtraction of the MPI contribution, a signal of the jet-induced diffusion wake similar to the $\gamma$-hadron correlation at RHIC \cite{Chen:2017zte}.

\noindent{\it 4. Enhancing diffusion wake with 2D jet tomography}:
In the above CoLBT-hydro results and experimental data,  $Z/\gamma$-hadron correlations are averaged over the initial transverse position and direction of the $Z/\gamma$-jets. Such an average smears out the angular structure of jet-induced medium response present in the $Z/\gamma$-hadron correlations when initial position and direction of $Z/\gamma$-jets \cite{Li:2010ts,Tachibana:2020mtb} are fixed. To overcome this smearing, we propose to utilize the longitudinal \cite{Zhang:2007ja,Zhang:2009rn} and the recently developed transverse gradient jet tomography \cite{He:2020iow} to localize the initial transverse positions of $Z/\gamma$-jets with given $Z/\gamma$ (or jet) direction relative to the event plane. According to the gradient tomography \cite{He:2020iow}, $Z/\gamma$-jet production points can be localized by the value of the transverse asymmetry,
\begin{equation}
    A_{\vec n}=\frac{\int d\phi [(dN^{h}/d\phi)_{\phi-\phi_n>0}-(dN^{h}/d\phi)_{\phi-\phi_n<0}]}{\int d\phi dN^{h}/d\phi},
\end{equation}
in each event with respect to a fixed jet plane $\vec n$ determined by $Z/\gamma$ (or the leading jet) and the beam.

\begin{figure}
\vspace{-0.8cm}
\centerline{\includegraphics[scale=0.58]{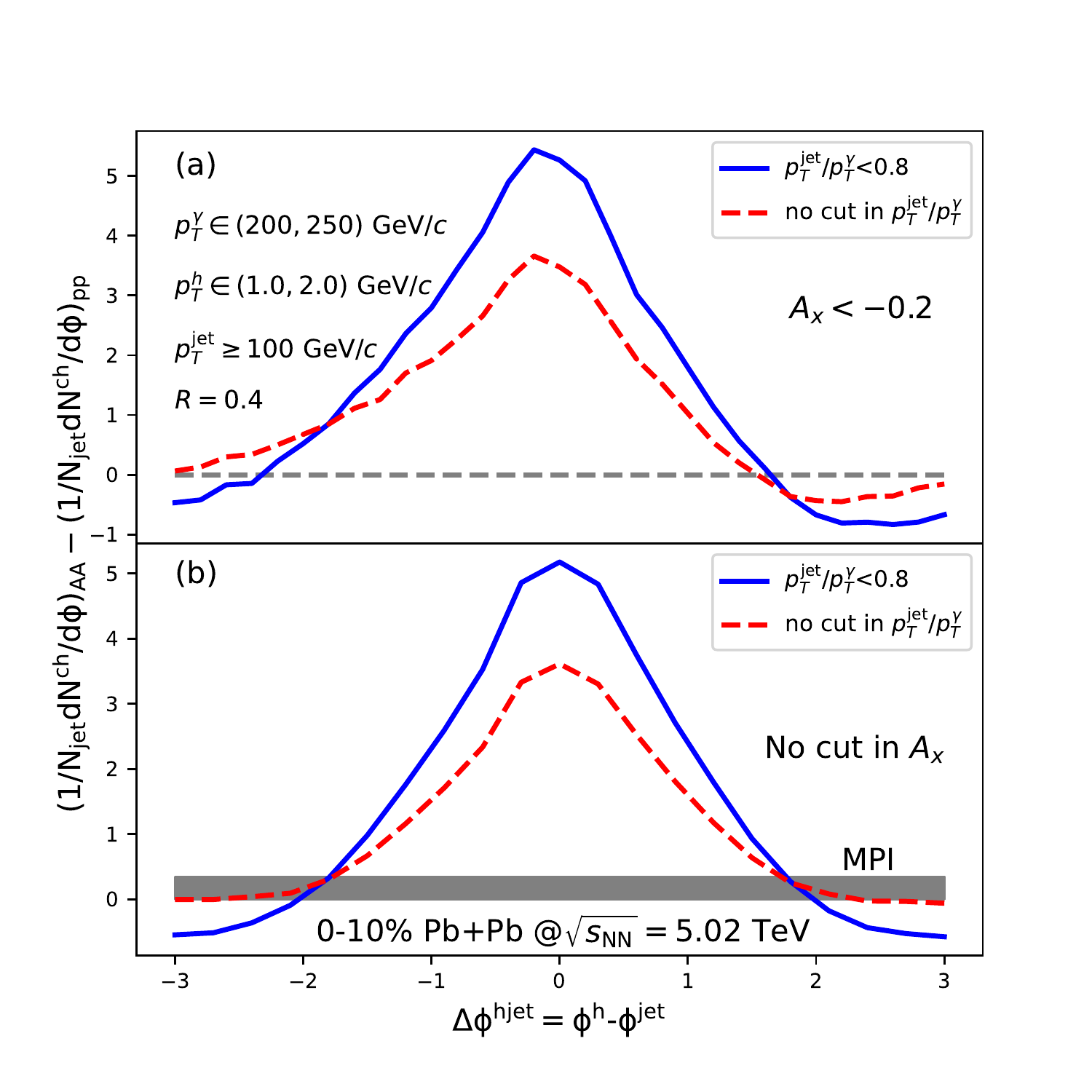}}\vspace{-0.5cm}
\caption{Difference in $\gamma$-hadron yields as a function of $\Delta\phi^{h\rm jet}$ between 0-10\% Pb+Pb and p+p collisions at $\sqrt{s}_{\rm NN}=5.02$ TeV for $p_T^h=1-2$ GeV/$c$, $p_T^\gamma=200-250$ GeV/$c$ and $p_T^{\rm jet}\ge 100$ GeV/$c$ (a) with and (b) without transverse asymmetry $A_x<-0.2$ for $p_T^h>2$ GeV/$c$ hadrons, (blue) with and (red) without $\gamma$-jet momentum asymmetry $p_T^{\rm jet}/p_T^\gamma<0.8$, MPI contribution is indicated by the gray band.}
\label{fig-asy}
\end{figure}

Shown in Fig.~\ref{fig-asy} (dashed lines) are medium modifications of soft hadron yields ($p_T^h=1-2$ GeV/$c$) in 0-10\% Pb+Pb collisions at $\sqrt{s}_{\rm NN}=5.02$ TeV as a function of $\Delta\phi^{h{\rm jet}}=\phi^h-\phi^{\rm jet}$ in $\gamma$-jet events (a) with and (b) without the transverse asymmetry $A_x<-0.2$ for $p_T^h>2$ GeV/$c$ hadrons with respect to the jet plane which is chosen to be perpendicular to the event plane here.
%(with $p_T^\gamma\in(100-150)$ GeV/$c$ and jet cone-size $R=0.4$) 
According to the gradient tomography, the initial transverse positions of $\gamma$-jets in events with the transverse asymmetry $A_x<-0.2$ are biased toward the upper half of the transverse plane and propagate tangentially as illustrated in Fig.~\ref{fig-machcone}. The jet-induced Mach-cone-like excitation in this configuration is more pronounced in the denser region of the medium leading to more enhancement of soft hadrons in $\Delta\phi^{h{\rm jet}}<0$ region. In the meantime, the diffusion wake is driven toward the less dense region so that the depletion of soft hadrons is deeper in $\Delta\phi^{h{\rm jet}}>0$ region. These asymmetric features and the diffusion wake are clearly seen  in the CoLBT-hydro results [Fig.~\ref{fig-asy}(a)] even without subtraction of the MPI background. In addition, one can select the value of the $\gamma$-jet momentum asymmetry $p_T^{\rm jet}/p_T^\gamma<0.8$, which biases toward longer jet propagation length \cite{Zhang:2007ja,Zhang:2009rn}. This will further enhance soft hadrons from the Mach-cone-like excitation in the jet direction as well as soft hadron depletion in the $\gamma$ direction due to the diffusion wake (solid lines), even without transverse asymmetry restriction [Fig.~\ref{fig-asy}(b)].  Subtraction of the MPI background (grey band) will further enhance the depletion of soft hadrons due to the diffusion wake.

\noindent{\it 5. Summary and discussions}:
We have investigated $Z/\gamma$-hadron correlation in high-energy heavy-ion collisions within the CoLBT-hydro model in search for the signal of jet-induced diffusion wake. The calculated hadron spectra with a $Z$-trigger show suppression of leading hadrons due to jet quenching and enhancement of soft hadrons due to the jet-induced Mach-cone-like excitation and medium induced soft gluon radiation, in good agreement with recent ATLAS and CMS data. The low $p_T^h$ $Z$-hadron correlation in azimuthal angle shows an enhanced and broadened peak in the jet direction as well as an enhancement in the $Z$ direction, which we identify as due to medium modification of partons from MPI. We have devised a procedure for the MPI background subtraction after which the effect of the diffusion wake in the $Z$ direction becomes visible as in the $\gamma$-hadron correlation at RHIC where MPI is negligible. We further apply the transverse gradient and longitudinal jet tomography technique to select events with localized initial $Z/\gamma$-jet productions. $Z/\gamma$-hadron correlations from these events are shown to have a clear signal of the diffusion wake even without the MPI subtraction. Future experimental measurements of the $Z/\gamma$-hadron correlation and the effect of the diffusion wake with this 2D jet tomography can provide further constraints on the bulk properties of the QGP in high-energy heavy-ion collisions.

{\it Acknowledgements}: We thank T.~Luo for discussions. This work is supported in part by %Guangdong Major Project of Basic and Applied Basic Research No. 2020B0301030008, 
NSFC under Grant Nos. 11935007, 11221504, 11861131009, 11890714 and 12075098, by DOE under Contract No. DE-AC02-05CH11231, by NSF under Grant No. ACI-1550228 within the JETSCAPE and OAC-2004571 within the X-SCAPE Collaboration, by the Fundamental Research Funds for Central Universities in China and by the UCB-CCNU Collaboration Grant. Computations are performed at Green Cube/GSI and NSC3/CCNU.

\bibliography{ref}

\end{document}